\documentstyle [aps,prc,preprint,rotating] {revtex}
\include{psfig}
\begin {document}
\date{\today}
\title {Quasi-gaussian fixed points and factorial cumulants in nuclear multifragmentation}
\author {D.~Lacroix$^1$\footnote{Service militaire actif, 
Scientifique du Contingent.}, R.~Peschanski$^2$. \\
 \medskip
(1) CEA, DAPNIA/SPhN CE-Saclay, 
91191 Gif-Sur-Yvette Cedex, France.\\
(2) CEA, Service de Physique Th\'eorique, CE-Saclay, 
91191 Gif-Sur-Yvette Cedex, France.\\}
\maketitle
\begin{abstract} 
We re-analyze the conditions for the phenomenon of intermittency (self-similar fluctuations) to occur in models of multifragmentation. Analyzing two different mechanisms, the bond-percolation and the ERW (Elattari, Richert and Wagner) statistical fragmentation models, we point out a common quasi-gaussian shape of the total multiplicity distribution in the critical range. The fixed-point property is also observed for the multiplicity of the second bin. Fluctuations are studied using scaled factorial cumulants instead of scaled factorial moments. The second-order cumulant displays the intermittency signal while higher order cumulants are equal to zero, revealing a large information redundancy in scaled factorial moments. A practical criterion is proposed to identify the gaussian feature of light-fragment production, distinguishing between a self-similarity  mechanism (ERW) and the superposition of independent sources (percolation). 
\end{abstract}
\begin{flushright}
{\bf DAPNIA/SPhN-96-17, SPhT T96/067}\\
Pacs: 24.60.Ky 25.70.Pq
\end{flushright}
\newpage

{\bf 1. Two classes of multifragmentation mechanisms: percolation and ERW models}

\medskip

In heavy-ion collisions, it is known that heavy excited systems are created and 
break into many lighter fragments of different sizes. This phenomenon is called multifragmentation.
It occurs when the 
energy deposited in the system is sufficient to develop instabilities but not 
enough to totally evaporate the system into nucleons. A statistical analysis of the average spectrum  
is consistent with a power-law dependence of the fragment-size distribution. Some interesting interpretations 
of multifragmentation based on a phase transition in nuclear matter have been
motivated by the experimental form of the fragment-size distribution\cite{Camp1,gro},
compatible with a power-law.
Indeed, several critical phenomena,
such as for instance percolation and liquid-gas phase transition exhibit a similar
power-law of the fragment-size distribution at the critical point\cite{stan,stau}. 
In these systems, the r\^ole of fragments is played 
by connected clusters of a given phase inside the other phase and the size distribution of
 clusters is proven to take the general form:
\begin{eqnarray}
N(s,\varepsilon)  \sim  s^{-\tau}\!\cdot f(s\ {\varepsilon}^{\sigma})\hspace{3cm}\tau, \sigma > 0 ,
\end{eqnarray}
where $s$ is the cluster size and $\varepsilon$ characterizes the distance from the critical 
point. In a thermal phase transition $\varepsilon=T-T_c$ with $T_c$ the critical
temperature. In a bond-percolation  
model $\varepsilon$ = $r_c-r$, with $r$ the probability 
for the bond between two neighbouring sites to be 
broken and $r_c$ its critical value\footnote{ In multifragmentation problems, it
is more convenient to use the variable
$r=1-q$, where $q$ is the usual bond probability\cite{stau}.}. The genuine phase transition is well-known 
to only occur when the system 
is infinite. However finite systems show specific features related to the continuous
limit, including finite-size corrections\cite{stan,stau}.
In multifragmenting systems,
the parameter $\varepsilon$ is not well identified. It could be associated
with the excitation energy deposited into the system by the collision. 

One typical property of phase transitions is the existence of large, scale 
invariant fluctuations near the critical point\cite{stan}.
In order to extract a signal from fluctuation patterns, one has proposed\cite{plo1} in nuclear physics
a method based on factorial
moments. This method was originally designed\cite{pes1} for the rapidity spectra of ultra-relativistic
multiparticle production but can be extended to other problems. 
Dividing the phase space into $M$ bins of size $\delta$, the factorial moment of order $p$ can be 
defined\footnote{We choose here the same definition as in Ref.\cite{plo1}.} by 
\begin{eqnarray}
F_p(\delta) =  \frac{\sum_{s=1}^{M}{<\!n_s(n_s-1)\cdots (n_s-p+1)\!>}}{\sum_{s=1}^{M}{{<\!n_s\!>}^p}} ,
\end{eqnarray}
\medskip
where $n_s$ is the number of "objects" in the $s^{th}$ bin and the average is taken over the 
set of events. In high-energy physics, "Objects" are particles distributed as a function 
of the rapidity variable. In nuclear physics, they are fragments distributed as a function
of the size. 

In particle physics, the interest of factorial moments has been to deconvolute the statistical 
fluctuations due to the limited number of "objects" observed by event. It has
revealed the existence of self-similar fluctuations, called {\it intermittency}, that is:

\begin{eqnarray}
F_p(\delta) \propto {\delta}^{-\alpha_p} ,
\end{eqnarray}
where $\alpha_p$ is the so-called intermittency exponent.  
For event-by-event
fluctuations of multifragmentation spectra\cite{plo1}, one considers 
$M$ bins of size $\delta$=$\frac{A}{M}$ along 
the fragment-size axis ($A$ being the mass of the excited system).
When $M$ is varied, a behaviour like (3) was observed in the fragment-size distribution 
of $^{197}_{79}$Au$_{118}$ nucleus 
multifragmenting in a nuclear emulsion\cite{fre}. The signal was compared with that 
obtained from 
a site-bond percolation model in a network of size 6$^3$, showing a similar behaviour at 
the critical point. The behaviour (3) was thus interpreted as a phase-transition signal.

However the application of the method to the fragment-size distribution
is less straightforward than for rapidity spectra of particles. Indeed, the expression (2) is 
dominated by the first bin, i.e. the one containing the lightest fragments. In particular, no fragment
of mass greater than $A/p$ can contribute to $F_p$. This introduces an 
important bias in the analysis, and has to be properly taken into account. 

On a less technical ground, several remarks have made the phase-transition
interpretations doubtful\cite{Camp2}. In percolation models, the signal disappears when the size of the system 
goes to infinity. Moreover, $F_p(\delta)$ seems to strongly depend 
on the shape of the total multiplicity distribution. In particular, when selecting events with the same
multiplicity, the signal disappears.

The interpretation is made more puzzling when considering 
a different class of multifragmentation mechanisms which leads 
to an intermittency behaviour without relying on a phase transition:
the ERW model\cite{ell}. Though there is no quantitative multifragmentation
model using this mechanism, its simplicity has made it a fruitful toy model
for understanding the mechanisms behind intermittency. 
In this case, the rule is to create fragments
of different sizes with a Monte-Carlo procedure as follows. The average fragment-size distribution
is constrained to be a power-law
distribution $P(s)$ $\sim$ $s^{-\tau}$, $\tau$ being a parameter of the model.
We note $A$ the total mass. An event is specified by a set of random numbers
$\eta_1,\eta_2,\cdots,\eta_n$ (with $0\leq\eta_i\leq 1$), which defines
a set of successive binary fragmentations (see Fig. 1). $\eta_1$ defines a first fragmentation into two 
fragments of mass $a_1$ and $A-a_1$, where $a_1$, an inert fragment present in the final
event, is implicitely defined by 
\begin{eqnarray}
\frac{\sum_{s=1}^{a_1-1}{s^{-\tau}}}{\sum_{s=1}^{A}{s^{-\tau}}}   <  \eta_1  \leq  \frac{\sum_{s=1}^{a_1}{s^{-\tau}}}{\sum_{s=1}^{A}{s^{-\tau}}} .  
\end{eqnarray}
The event is completed by repeating the operation with the mass $A-a_1$ (or $A-\sum_{i=1}^{k}{a_i}$ after $k$ 
generations) until the total mass is exhausted. This model has many interesting properties\cite{ell},
in particular, an intermittent
signal can be seen for $1.8< \tau <2.0$. 

Using the two different classes of mechanisms, percolation and ERW models 
we will re-examine the problem of interpreting intermittency 
by asking the following questions:

(i) Is there a common  scale-invariant mechanism behind 
the different realisations of intermittency?

(ii) Are there statistical tools which could be more suitable than
factorial moments for the study of multifragmentation spectra?

(iii) Can we distinguish between different mechanisms?

The plan of our study is as follows.        
In section 2, we show in both models
the existence of fixed points, where the shape of the multiplicity is stable
when the size of the system goes 
to infinity. This shape is {\it quasi-gaussian} for the total multiplicity and 
bin-size invariant but not gaussian when selecting the second bin.
In section 3, we show that {\it factorial cumulants}\cite{car}, 
instead of moments, are best suitable for the study of gaussian fluctuations
of the multiplicity spectra.
In the final section, we propose a criterion to distinguish between models,
together with a physical interpretation of the different mechanisms in action.

{\bf 2. Quasi-gaussian fixed points} 

Let us introduce well-known coefficients which characterize the shape of a statistical
distribution. For a given 
variable $m$, these are defined by
\begin{eqnarray}
\gamma_i = \frac {<{(m-<m>)}^{(i+2)}>} {{<(m-<m>)^2>}^{(\frac{i+2}{2})}}
\end{eqnarray}
In particular, we will consider 
the  skewness ($\gamma_1$) and sharpness ($\gamma_2$) coefficients\footnote{
Here, $\gamma_2$ should not be confused
with Campi's notation\cite{Camp1}.}.
For a gaussian probability distribution $\gamma_1=0$ and $\gamma_2=3$. The number $\gamma_1$ (resp. $\gamma_2$)
describes the main assymetric (resp. symetric) variation with respect to the gaussian. 
Interestingly enough, we have found fixed
points in the parameter space, where the values of $\gamma_1$, $\gamma_2$ evaluated 
for the total multiplicity nearly reach gaussian
values. We call them {\it quasi-gaussian} fixed points.

We have performed a systematic analysis of the 
multiplicity distributions in the ERW (Fig. 2) and percolation (Fig. 3) models.
Using ERW, we have first
plotted the values taken by $\gamma_1$ and $\gamma_2$ for the total multiplicity distribution as a function
of $\tau$ (see Fig. 2-a) for different values of $A$ ($50 \leq A \leq 500$). 
Each of the points reported on these figures is calculated with statistics of
about 10$^5$ events. We note the existence of a value $\tau \sim 1.8$ where 
all curves intersect.
Interestingly enough, this value is in the range where the 
intermittency behaviour has been noticed\cite{ell}. 
The values of $\gamma_1$ and $\gamma_2$ at the "fixed point" are 
close respectively  to $0$ and $3$ but nevertheless slightly different. The 
total multiplicity distribution is thus quasi-gaussian. We also note the existence
of a minimum
for $\gamma_2$ near $\tau$ $\sim$ $2$ which confirms
that the intermittent signal occurs when fluctuations are maximal\footnote{
 When $\gamma_2 > 3$, the distribution is sharper than a gaussian, 
and the fluctuations are weaker. On the contrary, the fluctuations are larger
when $\gamma_2 < 3$ (for instance $\gamma_2$ is equal to 1.8 for a
uniform distribution). In our study, we found a minimum of $\gamma_2$ near $2.5$.}. 
This study has revealed the existence of a scale-invariant property of the
total multiplicity distribution.

In order to understand its connection with the intermittency signal, an analysis at fixed total 
size but varying the bin size appears useful.
We thus studied in more detail the multiplicity distribution in the second bin 
$n_2(\delta)$ (Fig. 2-b). The results are shown for $\gamma_1$ and $\gamma_2$ as a function of $\tau$. 
The curves correspond to varying values of the bin size ($4 \leq \delta \leq 64$) 
with a fixed value of the system size $A=128$. 
There is again a clear evidence for a "fixed point". 
We have also verified that this fixed point remains when the system size increases. We note that 
the corresponding value of $\tau$ is similar but slightly different than that of Fig. 2-a. Nevertheless,
it remains in the region where the intermittent behaviour is noticed.
The same analysis
for the first bin distribution $n_1(\delta)$ is not relevant
because the values of $\gamma_1$ and $\gamma_2$ are strongly dominated by the distribution 
of fragments of mass one. In order to study the scale-invariance properties as a function
of the bin size $\delta$, it is thus necessary to avoid as much as possible the statistical
bias due to the dominance of mass-one fragments. Moreover, there could be good experimental
reasons to avoid the contribution of lightest fragments\cite{che} which   
can be contaminated
by other processes than nuclear multifragmentation (e.g. pre-equilibrium, secondary emission, etc...).

The same analysis was performed with the finite-size bond percolation model (Fig. 3).
One considers a 3-dimensional cubic lattice with a probability $r$ of breaking a bond.
The critical value  of the bond-breaking probability $r_c$ is around $0.76$ ($0.7512$ in the continuous
limit\cite{stau}). We have first plotted the values of $\gamma_1$ and $\gamma_2$ for the total multiplicity
distribution as a function of $r$ for various sizes of the system ($4^3 \leq A \leq 10^3$),
see Fig. 3-a. Each point corresponds to the same statistics ($10^5$ events) as for the previous study.
There is evidence for a fixed point in $\gamma_1$ near $r_c$. 
The fixed-point behaviour is less clear for the sharpness 
parameter $\gamma_2$. The values of $\gamma_1$ and $\gamma_2$ remain respectively near $0$
and $3$ confirming the existence of a quasi-gaussian shape in the critical region of the parameter.
In the same region, intermittency has been observed\cite{plo1}. In the
second bin, the analysis of $\gamma_1$ and $\gamma_2$ (Fig. 3-b) clearly indicates
the existence of a fixed point in the same parameter range.

The strong similarities between the two models and the location of 
the fixed point in the same parameter region where intermittency occurs,
calls for a deeper analysis of intermittency. Indeed, the quasi-gaussian
features of the total multiplicity distribution 
require a specific treatment of intermittency which goes beyond the use   
of factorial moments.  

{\bf 3. Factorial cumulants}

The existence of quasi-gaussian distributions leads us to introduce new statistical tools
for multifragmentation, namely the {\it factorial cumulants}. In particle physics, 
they have been introduced\cite{car} in order to combine the elimination
of the statistical noise using the factorial form with the well-known property of cumulants. Cumulants are a-priori able to  
disentangle genuine higher-order
correlations from combinations of lower-order correlations which appear in factorial moments. In particular, gaussian fluctuations
are governed by 2-body correlations implying the vanishing of cumulants of order
greater than two. Let us apply this tool to nuclear multifragmentation 
taking into account the specific normalisation used in formula (2).
We first introduce the generating function of unscaled factorial moments
\begin{eqnarray}
G(\lambda)=\sum_{s=1}^{M}{<\!\lambda^{n_s}\!>} ,
\end{eqnarray}
where $\lambda$ is an arbitrary parameter and the average $<\cdot>$ is performed over the set of events.
For instance, formula (2) can be identically written as
\begin{eqnarray}
F_p(\delta)&=&\left. \frac{\left( \frac{\partial^p G(\lambda)}{\partial\lambda^p}\right)}{\left( \frac{\partial G(\lambda)}{\partial\lambda}\right)^p} \right|_{\lambda=1}.
\end{eqnarray}
Using the logarithm of the generating function, the general expression of the scaled factorial cumulants can be written
\begin{eqnarray}
K_p(\delta)&=&\left. \frac{\left( \frac{\partial^p \ln G(\lambda)}{\partial\lambda^p}\right)}{\left( \frac{\partial G(\lambda)}{\partial\lambda}\right)^p} \right|_{\lambda=1}.
\end{eqnarray}
For instance, the first cumulants read
\begin{eqnarray}
K_2(\delta)&=&\frac{\sum_{s=1}^{M}{\left( <\!n_s(n_s-1)\!>-<\!n_s\!>^2 \right)}}{\sum_{s=1}^{M}{<\!n_s\!>^2}} , \nonumber  \\ 
K_3(\delta)&=&\frac{\sum_{s=1}^{M}{\left( <\!n_s(n_s-1)(n_s-2)\!>-3<\!n_s(n_s-1)\!><\!n_s\!>+2<\!n_s\!>^3  \right)}}{\sum_{s=1}^{M}{<\!n_s\!>^3}} , \nonumber \\
K_4(\delta)&=&\frac{\sum_{s=1}^{M}{\left( <\!n_s(n_s-1)(n_s-2)(n_s-3)\!>-4<\!n_s(n_s-1)(n_s-2)\!><\!n_s\!> \right)}}{\sum_{s=1}^{M}{<\!n_s\!>^4}} , \nonumber \\
           &+&\frac{\sum_{s=1}^{M}{\left( -3<\!n_s(n_s-1)\!>^2+12<\!n_s(n_s-1)\!><\!n_s\!>^2-6<\!n_s\!>^4 \right)}}{\sum_{s=1}^{M}{<\!n_s\!>^4}} . 
\end{eqnarray}
It is interesting to note that the choice of normalisation (2)
implies that the scaled factorial cumulants cannot in general be expressed 
as combinations of scaled factorial moments. If however, the first bin 
dominates the evaluation of moments, one obtains the following approximate
relations:
\begin{eqnarray}
K_2(\delta)&\simeq&F_2(\delta)-1 ,\nonumber \\ 
K_3(\delta)&\simeq&F_3(\delta)-3\cdot F_2(\delta)+2 \nonumber ,\\ 
K_4(\delta)&\simeq&F_4(\delta)-4\cdot F_3(\delta)-3\cdot {(F_2(\delta))}^2+12\cdot F_2(\delta)-6 . 
\end{eqnarray}  
Note that the direct determination of cumulants using formula (8) could be useful to avoid the errors on the combinations of factorial moments. 

Let us apply our formalism to the ERW and percolation models.
We have first computed the factorial moments using definition (1), see Fig. 4-a and 4-b. The result 
obtained with the ERW model is displayed in Fig 4-a with a total mass $A=128$ using the critical value $\tau=1.8$.
The figure 4-b corresponds to the same observables for the bond-percolation model with 
$A=6^3$ and $r \simeq 0.75=r_c$. The curves  reproduce (for slightly different parameter values) the results of
refs.\cite{plo1} and \cite{ell}. Note that, for completion, we have also reported
the ERW factorial moments for the value $\tau = 2.3$ 
corresponding to the slope observed in the critical region of the bond-percolation model.
The results exhibit a nearly linear increase as a function of the bin size which is typical 
of an intermittent behaviour. 

We have calculated the factorial cumulants 
corresponding to formula (9) for the ERW (Fig. 4-c) and percolation (Fig. 4-d) models. Cleary, only the second cumulant $K_2(\delta)$ is significantly different from 0. The higher 
order factorial cumulants are nearly zero for all bin size $\delta$. Note that due to the smallness and the 
change of sign of cumulants, it is not convenient to plot their logarithm. Indeed, the cumulants obtained for both models correspond to strong cancellations between the large and positive scaled factorial moments.
This confirms that there is a redundancy of information in factorial moments which is solved by using cumulants.

The figures 4-c and 4-d show a clear indication of a rise of $K_2(\delta)$ especially for the percolation
model. When compared to Fig. 5, where scaled factorial cumulants are displayed for the bond-percolation model
outside the critical
region, the behaviour of $K_2(\delta)$ is revealing intermittent fluctuations. 
Indeed, an intermittent singularity in factorial moments should
also appear in the second cumulant since it cannot be absorbed by the lower order contributions. For small values of $K_2(\delta)$ the first relation (10) gives 
\begin{eqnarray}
K_2(\delta) \ \simeq \ \log(F_2(\delta)) \ \propto \ -\alpha_p \ \log(\delta) .
\end{eqnarray}
Such a behaviour is observed in Fig. 4. It is interesting to note that the 
dominance of second-order cumulants is valid for all values of the parameter $r$ in the percolation model.  

Considering the abovementionned properties of scaled factorial cumulants for multifragmentation mechanisms,
a series of comments are in order:

(i) The scaled factorial moments of multifragmentation models 
seem to contain redundant information, since only the second cumulant is significantly different 
from zero. In the framework of the approximation (10), 
which is valid since the first bin is largely
dominating, it means that the scaled factorial moments can all be expressed as functions of $K_2(\delta)$.

(ii) In the percolation model, the cumulant analysis indicates 
that the same property still holds outside the critical region, see Fig. 5. However,
it is only around the critical region $r \sim 0.75$ that one observes a rise of $K_2(\delta)$
specific of intermittency.

(iii) A qualitative difference exists between the behaviour of $K_2(\delta)$ for the ERW and the 
percolation models. They take very different values for the largest bin size ($\delta \equiv A$),
i.e. the second cumulants of the total multiplicity distributions are significantly different.

(iv) We also remarked that 
the slope of $K_2(\delta)$ decreases with increasing system size, similarly to what has been observed
for scaled factorial moments\cite{Camp2}. However, the order-of-magnitude difference  still remains between the values 
of $K_2(\delta)$ for the ERW and the percolation models.
\medskip

{\bf 4. Discussion and summary of results}

Our analysis of multifragmentation spectra using the comparison 
of two generic mechanisms, ERW and percolation, calls for a physical interpretation.
The existence of a scaling behaviour for the profile parameters $\gamma_1$, $\gamma_2$, being a common feature
of both models in their respective critical regime, points to the existence of a scale-invariant property of
multiplicity distributions. It is phenomenologically clear that this "fixed-point"   is related to the intermittency 
signal and thus to a certain type of {\it criticality} of the system. The criticality is explicit for percolation,
since it corresponds to a phase transition in the continuous limit, 
but is hidden in the formulation of the ERW mechanism.

Observing at the same time a quasi-gaussian shape of the total multiplicity and a
large dominance of $K_2(\delta)$ over higher order cumulants unravels a pronounced gaussian
feature of multifragmentation mechanisms when small-mass fragments dominate the observables. Two main 
interpretations may naturally explain these features. It may come either from the existence of a self-similar fragmentation mechanism
with a scale invariant gaussian fixed-point or from the superposition of many independent sources
of fragments leading to a gaussian distribution through the central-limit theorem. 
As we once more want to stress, the gaussian property is mainly inherent to the
production of lighter elements.

In order to distinguish between the two options, let us study the two parameters 
which define a gaussian distribution namely the distribution average $<m>$ and its variance $\sigma=<(m-<m>)^2>$.  
One expects\cite{meu}
 $\sigma/\!<\!n\!> \rightarrow cste$ for self-similar models and $\sigma/\!<\!n\!>$ $\propto$ $1/\!\sqrt{<\!n\!>}$
for the independent sources. Indeed, in this latter case the number of independent sources is expected to grow with the multiplicity. Then the central-limit theorem gives the quoted prediction for $\sigma/\!<\!n\!>$. In Fig. 6, we display the quantity $\sigma/\!<\!n\!>$ for each model
for both the total multiplicity distribution and the mass-one fragment distribution. For the total multiplicity, 
power-law fits give $\sigma/\!<\!n\!>$ $ \propto$ $ {<\!n\!>}^{-0.01}$ for the ERW model and 
$\sigma/\!<\!n\!>$ $\propto $ $ {<\!n\!>}^{-0.44}$ for the percolation model.
Interestingly enough, the fits are very similar for the mass-one fragments with a power dependance ${<\!n\!>}^{-0.49}$
for the percolation model very suggestive of a central-limit signature. Hence, the ERW is a self-similar
mechanism, while percolation is gaussian with a central-limit type for the production of light fragments.

Let us summarize our main results by proposing the following new tools of analysis for multifragmentation fragment distributions.

{\bf (i)} The shape parameters $\gamma_1$, $\gamma_2$ for the total multiplicity and for the second-bin multiplicity 
can be studied as a function 
of the system size, and for the second case as a function of the bin size. We expect 
a fixed-point behaviour for the critical region of multifragmentation.
Intermittency is expected to occur in the same region. Note that the second bin 
could be experimentally easier to study, since it avoids the ambiguity on the origin of lightest
fragments and can be used at fixed total-size by varying only the bin-size.

{\bf (ii)} The study of fluctuations using scaled factorial moments suffers from a large redundancy of information.
For the class of models under study, we suggest using instead the scaled factorial cumulants of fluctuations.

{\bf (iii)} The gaussian feature of the light-fragment distributions revealed both by the quasi-gaussian fixed point
property and the scaled factorial cumulants is an important characteristics to be studied in nuclear multifragmentation. Note that this gaussian feature has also been noticed in a different model\cite{gro2}, confirming the interest of studying this property in detail. Two mechanisms seem to give an alternative for explaining the gaussian features, namely either a self-similar fixed-point or a superposition of independent
emission sources for light fragments. They can be distinguished by   dependence of $\sigma/\!<\!n\!>$ with $<\!n\!>$ or
equivalently by the factorial cumulants for large bins.  As an application, we found that the ERW model is self-similar
while percolation probably leads to independent sources for the production of light fragments 
even outside the critical region. We expect the self-similar property to be shared by models based on a tree structure of multifragmentation\cite{gir,plo2}. In the same spirit we also expect models based on second-order phase transitions to follow the same trend governed by the central-limit theorem as the percolation model for the distribution of light fragments.

{\bf ACKNOWLEDGMENTS}
 
The authors thank Roland Dayras and Jean-Louis Meunier for many fruitful discussions and Bertrand Giraud for a careful critical reading of the manuscript. One of us (D.L.) thanks the SPHN
for its hospitality and physicists for their kindness.

\begin{figure}
\hspace{4.5cm}
\psfig{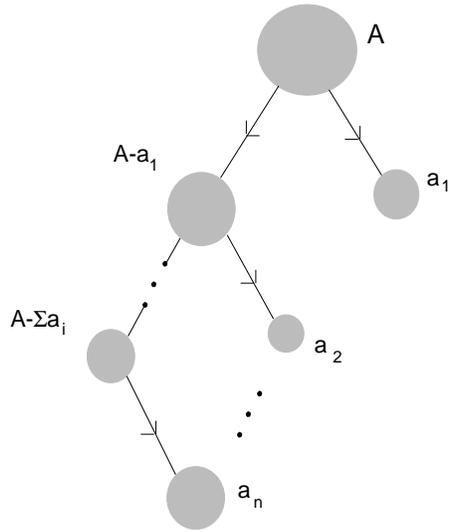}\\
\caption{ {\it Event generation in the ERW model}. The set of successive fragments
$a_1,a_2\cdots,a_n$ is defined by the set of 
random number $\eta_1,\eta_2,\cdots,\eta_n$ (see formula (4)). The
generation of random numbers is stopped when $\sum_{i=1}^{n} {a_i} > A$. }
\end{figure}

\newpage

\begin{figure}
\begin{tabbing}
\psfig{figure=g1mer.epsi,height=6.5cm} \= 
\psfig{figure=g2mer.epsi,height=6.5cm} \\
\put(0,0) {~~~~~~~~~~~~~~~~~~~~~~~~~~~~~~~~~~~~~~~~~~~~~~~~~~~~~~~~~~~~~~~~{a)}}\\
\psfig{figure=bin2g1erw.epsi,height=6.5cm} \=
\psfig{figure=bin2g2erw.epsi,height=6.5cm} \\
\put(0,0) {~~~~~~~~~~~~~~~~~~~~~~~~~~~~~~~~~~~~~~~~~~~~~~~~~~~~~~~~~~~~~~~~{b)}}\\
\end{tabbing}
\caption{ {\it Shape parameters $\gamma_1$, $\gamma_2$ of the multiplicity distributions for
the ERW model.} \\
a) Top: $\gamma_1$, $\gamma_2$ (total multiplicity 
distribution) as a function of 
$\tau$ for different system sizes ($A=50,100,200,300,400$ and $500$). 
b) Bottom: $\gamma_1$, $\gamma_2$ (second bin) as a function of 
$\tau$ for different values of the bin size $\delta$ ($A=128$).}
\end{figure}
\begin{figure}
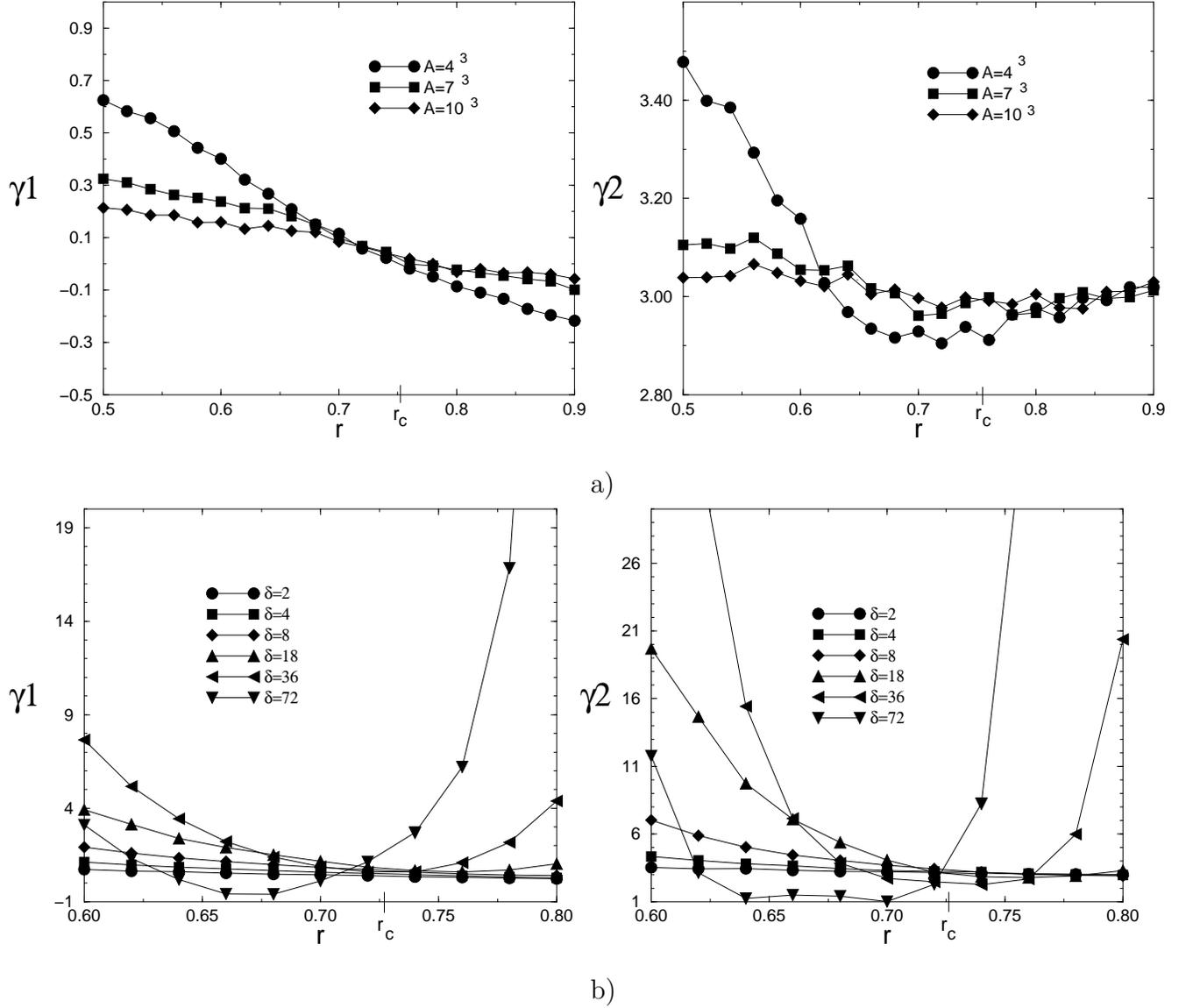

\centering
\begin{tabbing}
\psfig{figure=g1p.epsi,height=6.6cm} \=
\psfig{figure=g2p.epsi,height=6.6cm} \\
\put(0,0) {~~~~~~~~~~~~~~~~~~~~~~~~~~~~~~~~~~~~~~~~~~~~~~~~~~~~~~~~~~~~~~~~{a)}}\\
\psfig{figure=g1pb1.epsi,height=6.6cm} \=
\psfig{figure=g2pb2.epsi,height=6.6cm} \\
\put(0,0) {~~~~~~~~~~~~~~~~~~~~~~~~~~~~~~~~~~~~~~~~~~~~~~~~~~~~~~~~~~~~~~~~{b)}}\\
\end{tabbing}
\caption{ {\it Shape parameters $\gamma_1$, $\gamma_2$  for multiplicity 
distributions for
the bond-percolation model.} a) Top: $\gamma_1$, $\gamma_2$ (total multiplicity distribution) as a function of 
$r$ for different system sizes $A$. 
b) Bottom: $\gamma_1$, $\gamma_2$ (second bin) as a function of 
$r$ for different values of the bin size $\delta$ ($A=6^3$).}
\end{figure}
\begin{figure} \unitlength=1.0\textwidth%
\begin{flushleft}
\begin{tabular} {cc} 
{\makebox(0.49,.005)[0,0]{a)}}%
&{\makebox(0.49,.005)[0,0]{b)}}%
\end{tabular}
\begin{tabular}[t]{cccc}
\begin{sideways}\makebox[0.3\unitlength]{$log(F_p(\delta))$}%
\end{sideways}&\psfig{figure=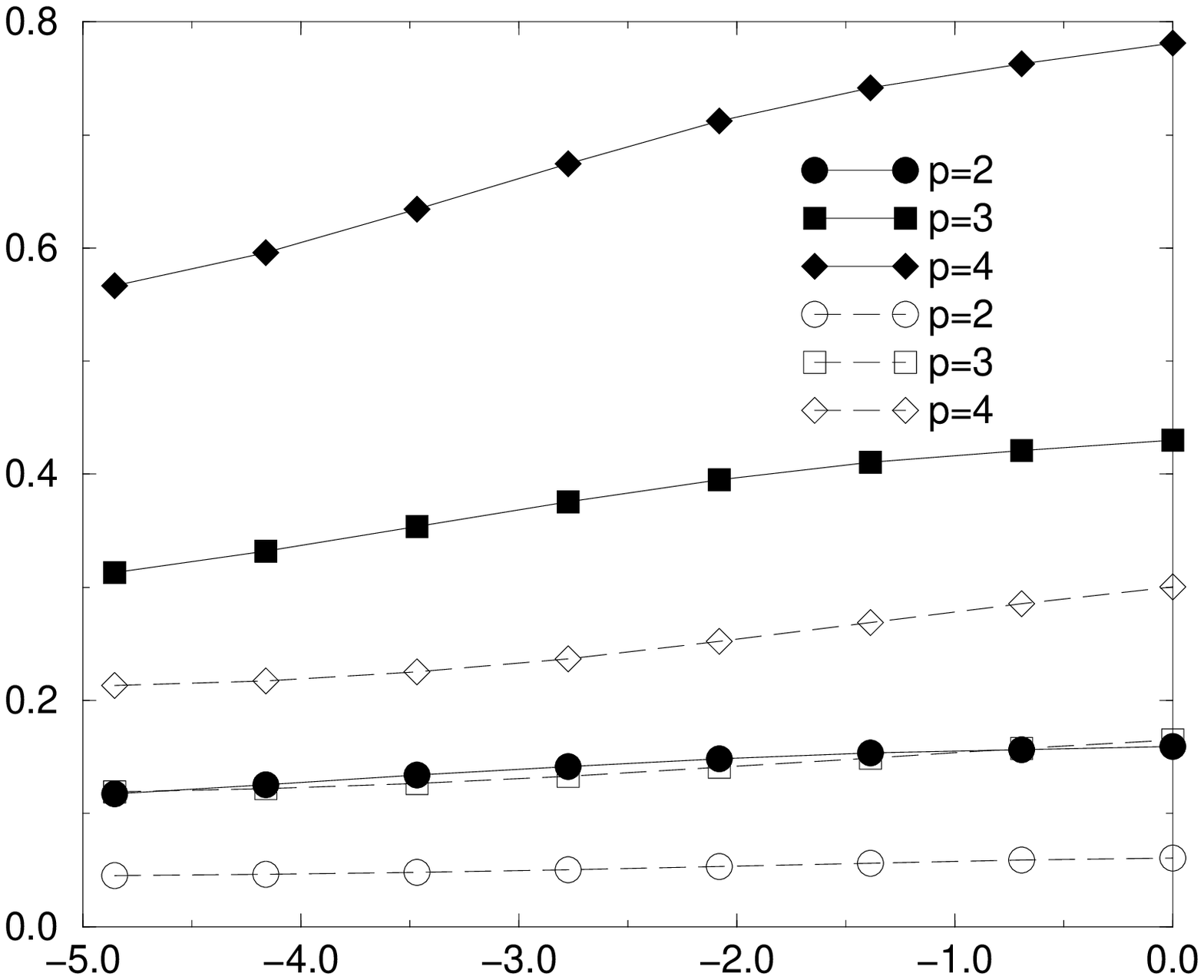,height=0.34\unitlength,width=0.38\unitlength}
&\begin{sideways}\makebox[0.3\unitlength]{$log(F_p(\delta))$}%
\end{sideways}&\psfig{figure=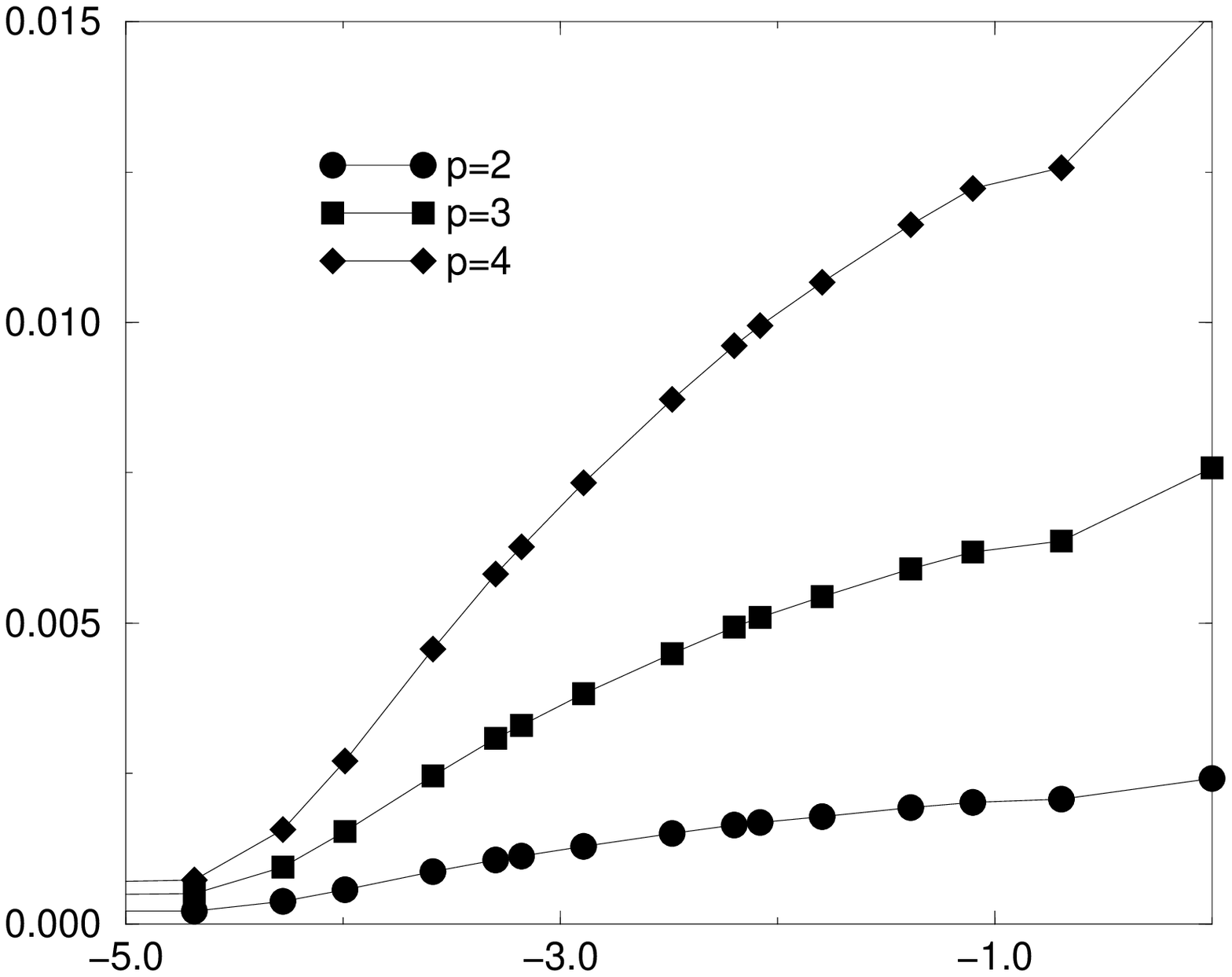,height=0.34\unitlength,width=0.38\unitlength}\\
&{\makebox(0.5,.005)[0,0]{$-log(\delta)$}}%
&\put(0,0){\makebox(0.5,.01){$-log(\delta)$}}%
\end{tabular}
\end{flushleft}
\begin{flushleft}
\begin{tabular} {cc} 
{\makebox(0.49,.005)[0,0]{c)}}%
&{\makebox(0.49,.005)[0,0]{d)}}%
\end{tabular}
\begin{tabular}[t]{cccc}
\begin{sideways}\makebox[0.3\unitlength]{$K_p(\delta)$}%
\end{sideways}&\psfig{figure=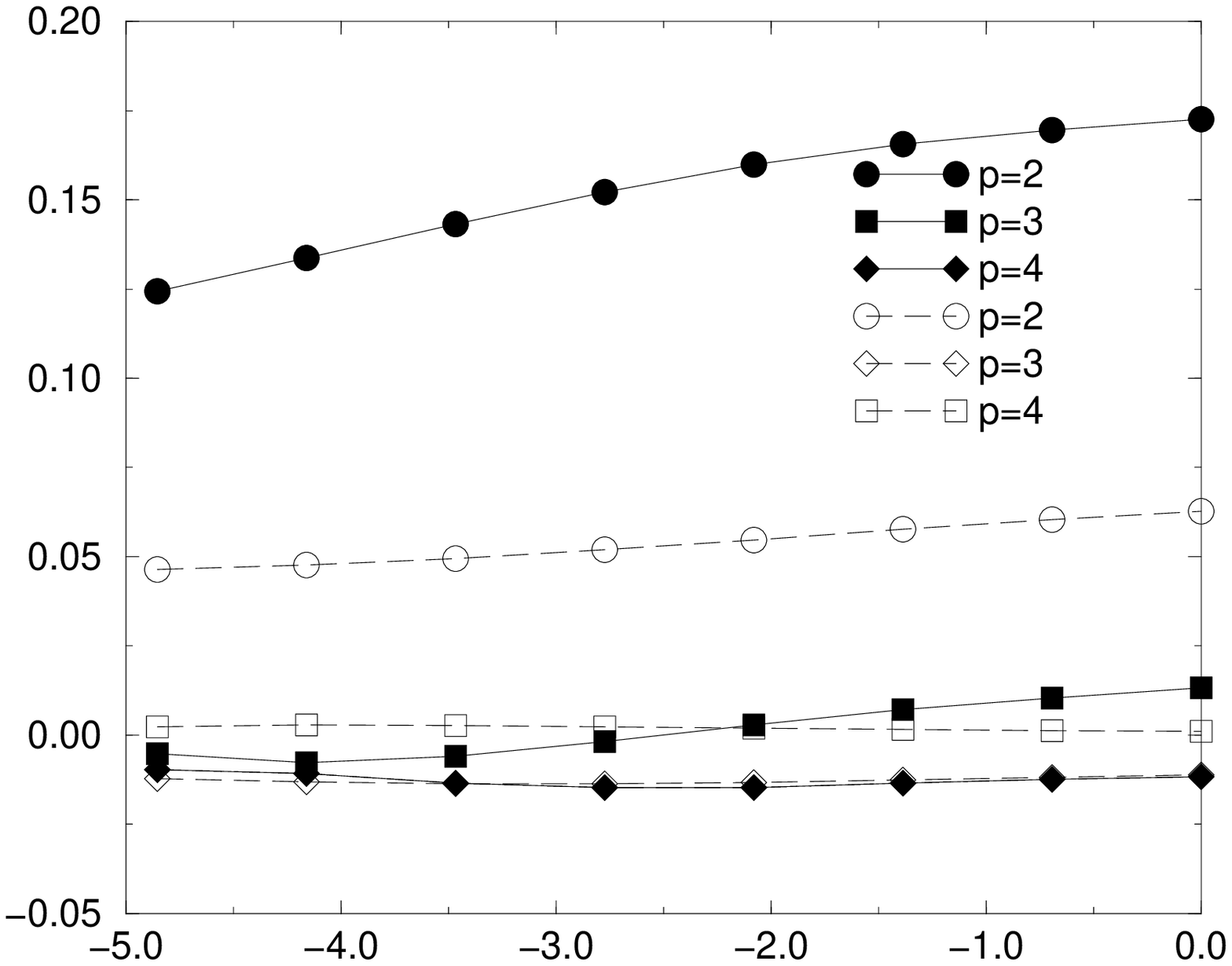,height=0.34\unitlength,width=0.38\unitlength}
&\begin{sideways}\makebox[0.3\unitlength]{$K_p(\delta)$}%
\end{sideways}&\psfig{figure=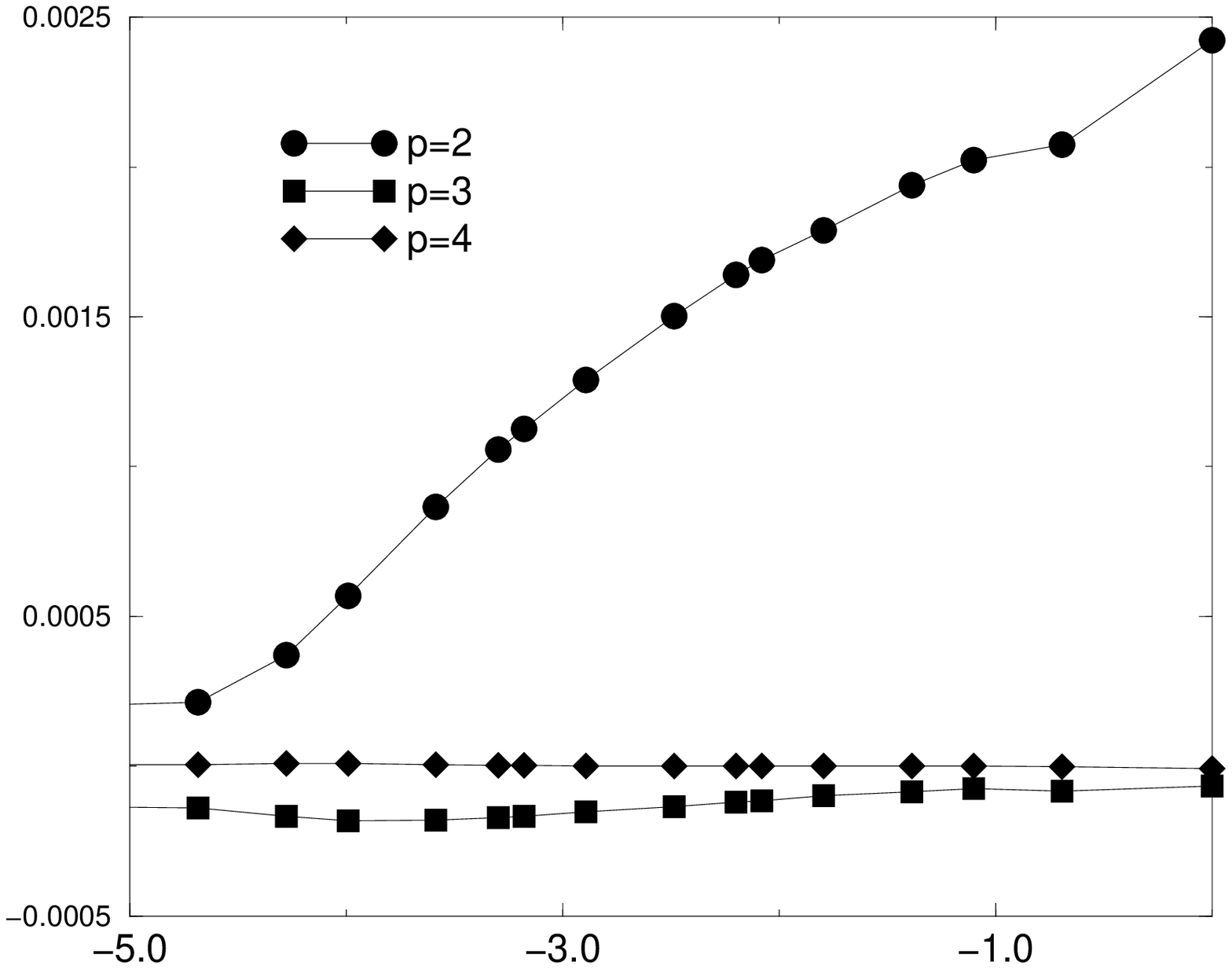,height=0.34\unitlength,width=0.38\unitlength}\\
&{\makebox(0.5,.005)[0,0]{$-log(\delta$)}}%
&\put(0,0){\makebox(0.5,.01){$-log(\delta$)}}%
\end{tabular}
\end{flushleft}
\caption{ {\it Scaled factorial moments $F_p(\delta)$ and cumulants $K_p(\delta)$.}
a) $\log(F_p(\delta))$ as a function of $-\log(\delta)$ in the ERW model   
for $A=128$, $\tau=1.8$ (solid line) and $\tau=2.3$ (dashed line).
b) $\log(F_p(\delta))$ as a function of $-\log(\delta)$ 
in the percolation model for $A= 6^3$ and $r=0.75$.
c) $K_p(\delta)$ vs $-\log(\delta)$ in the ERW model
for $A=128$ and $\tau=1.8$ (solid line) and $\tau=2.3$ (dashed line).
d) $K_p(\delta)$ vs $-\log(\delta)$ in the percolation model 
for $A=6^3$ and $r=0.75$. }
\end{figure}
\newpage
\begin{figure}
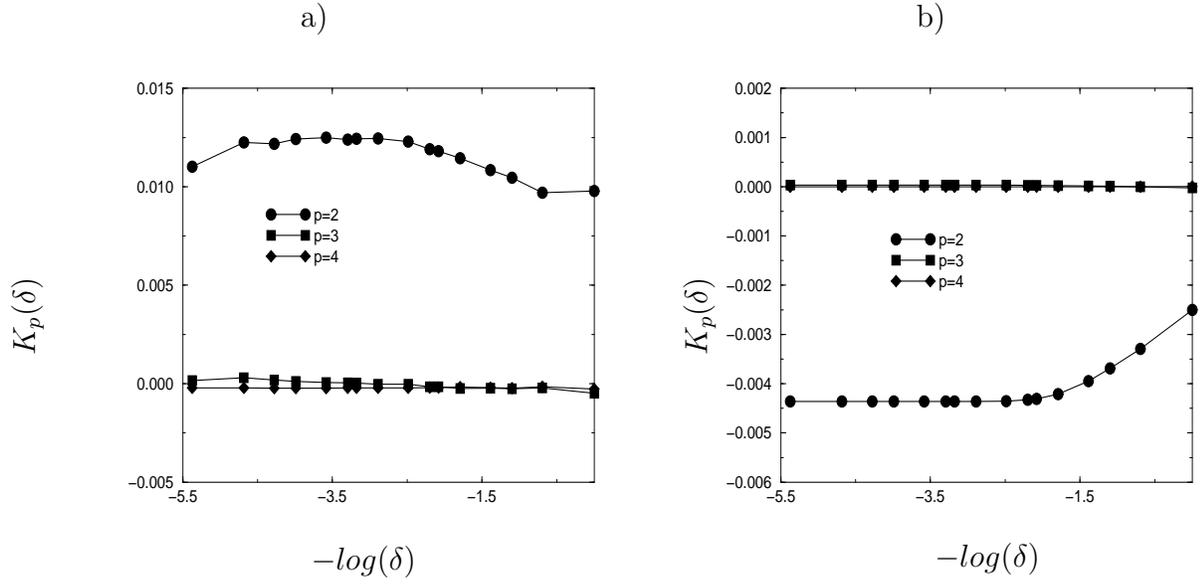
 \unitlength=1.0\textwidth%
\begin{flushleft}
\begin{tabular} {cc} 
{\makebox(0.49,.01)[0,0]{a)}}%
&{\makebox(0.49,.01)[0,0]{b)}}%
\end{tabular}
\end{flushleft}
\begin{tabular}[t]{cccc}
\begin{sideways}\makebox[0.3\unitlength]{$K_p(\delta)$}%
\end{sideways}&\psfig{figure=p10cum.epsi,height=0.34\unitlength,width=0.38\unitlength}
&\begin{sideways}\makebox[0.3\unitlength]{$K_p(\delta)$}%
\end{sideways}&\psfig{figure=p40cum.epsi,height=0.34\unitlength,width=0.38\unitlength}\\
&{\makebox(0.5,.005)[0,0]{$-log(\delta)$}}%
&\put(0,0){\makebox(0.5,.01){$-log(\delta)$}}
\\ 
\\
\end{tabular}
\caption{{\it Scaled factorial cumulants $K_p(\delta)$ outside the critical region for the bond-percolation
model.} a) $K_p(\delta)$ as a function of $\delta$
for $r=0.90$ and $A=6^3$. b) $K_p(\delta)$ 
for $r=0.60$ and $A=6^3$.} 
\end{figure}
\newpage
\begin{figure}
\psfig{figure=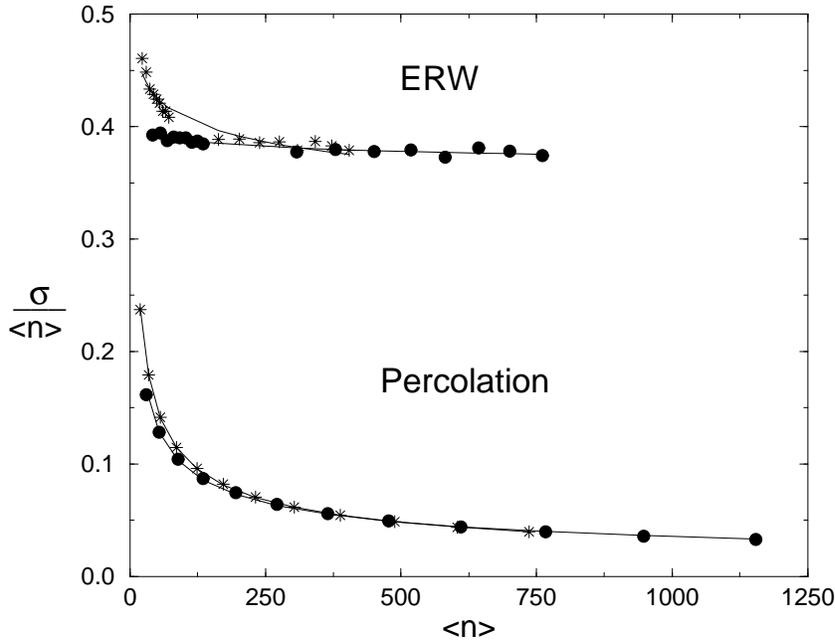,height=10cm}
\caption{ {\it The scaled variance $\sigma/\!<\!n\!>$ as a function of $<\!n\!>$.}
$\sigma/\!<\!n\!>$ is displayed as a function of the average multiplicity. Circles correpond
to the variance and average of the total multiplicity distribution; 
stars correspond to the variance and average multiplicity distribution of mass-one fragments.
For the ERW model with $\tau=1.8$, each point corresponds to a given total mass $A$ taken 
in the intervalle [100,10000]. Power-law fits are displayed by continuous lines, 
the powers are $-0.01$ (circles) and $-0.06$ (stars) in the ERW case.
For the bond-percolation model, $r=0.76$ and each point corresponds to a fixed value of 
$A$ ($A=3^3,4^3,\cdots,20^3$). The power-law fits are found respectively equal to $-0.44$ (circles)
and $-0.49$ (stars).}
\end{figure}
\end {document}